\documentclass[prb,twocolumn]{revtex4}%

\usepackage{graphicx}
\usepackage{amsmath}
\usepackage{ulem}
\usepackage{color}

\newcommand{\be}{\begin{equation}}
\newcommand{\ee}{\end{equation}}
\newcommand{\bea}{\begin{eqnarray*}}
\newcommand{\eea}{\end{eqnarray*}}
\newcommand{\bean}{\begin{eqnarray}}
\newcommand{\eean}{\end{eqnarray}}

\begin{document}

\draft
\title{\bf Environmental Breakdown of Topological Interface States in Armchair Graphene Nanoribbon Heterostructures}

\author{David M T Kuo}

\address{Department of Electrical Engineering and Department of Physics, National Central
University, Chungli, 32001 Taiwan}

\date{\today}

\begin{abstract}
We theoretically investigate the stability and transport
properties of topological interface states (IFs) in 9-7-9 and
15-13-15 armchair graphene nanoribbon heterostructures (AGNRHs)
laterally embedded in boron nitride (BN) sheets. Two
configurations, $n$-BNNR/AGNRH/$n$-BNNR and
$n$-BNNR/AGNRH/$n$-NBNR, corresponding to same-topology and
reverse-topology BN environments, are examined within a
tight-binding framework. Using a bulk boundary perturbation
approach, we show that in BNNR/AGNRH/BNNR the IFs are destroyed by
chirality breaking induced by symmetric BN environments at both
interfaces. In contrast, the IFs in the reverse-topology structure
remain robust against lateral interface interactions from BN
atoms. Transport calculations further demonstrate that the
surviving IFs in BNNR/AGNRH/NBNR exhibit the characteristic
behavior of topological double quantum dots, with an enhanced
interdot hopping strength compared with vacuum boundary
conditions. These results reveal that BN environments can either
suppress or reinforce topological interface states, depending
critically on the topology of the surrounding nanoribbons.
\end{abstract}

\maketitle

\section{Introduction}
Bottom-up synthesis techniques enable the fabrication of graphene
nanoribbons (GNRs) with atomic precision
[\onlinecite{Cai}--\onlinecite{DJRizzo}]. Compared with zigzag
GNRs (ZGNRs), armchair GNRs (AGNRs) exhibit tunable electronic
phases [\onlinecite{LlinasJP}], making them promising candidates
for nanoelectronic applications [\onlinecite{WangHM}]. More
complex GNR architectures, including 9-7-9 and 7-9-7 armchair
graphene nanoribbon heterostructures (AGNRHs)
[\onlinecite{DJRizzo}] and Janus GNR segments
[\onlinecite{SongST}], have also been experimentally realized.
Scanning tunneling microscopy measurements have confirmed the
presence of topological states (TSs) in these systems, such as end
states (ESs) and interface states (IFs) in 9-7-9 and 7-9-7 AGNRHs
[\onlinecite{DJRizzo}].

Although topological IFs in AGNRHs, such as 9-7-9 ribbons, have
been extensively studied in idealized or substrate-agnostic models
[\onlinecite{DJRizzo},\onlinecite{DavidKuo},\onlinecite{DavidKuo1}],
the role of the environment particularly lateral embedding
materials that break sublattice symmetry remains largely
unexplored. In realistic device configurations, AGNRHs are
inevitably coupled to surrounding materials, raising a critical
question: are topological interface states robust against such
environmental perturbations ? Here, we theoretically investigate a
9-7-9 AGNRH laterally embedded in boron nitride nanoribbons
(BNNRs), as illustrated in Fig.~1(a) and 1(b). Two distinct
configurations are considered, in which the upper and lower BNNRs
realize either the same-topology or reverse-topology scenario.
Owing to the difficulty of employing density functional theory
(DFT) to compute transmission coefficients in large
9-7-9-AGNRH/BNNR junctions
[\onlinecite{SonYW}--\onlinecite{NikitaVT}], we adopt a
tight-binding model combined with the Green function technique to
evaluate quantum transport in AGNRHs embedded in BN environments
[\onlinecite{QiaoZH},\onlinecite{JungJ}].

Surprisingly, we find that BNNR/AGNRH/NBNR structures with reverse
topology preserve the functionality of topological double quantum
dots (TDQDs), in which a single energy level associated with each
quantum dot remains well isolated from the bulk states. Moreover,
the electron hopping strength between the TDQDs is significantly
enhanced compared with the vacuum boundary condition. This
behavior is highly unusual when contrasted with conventional
semiconductor DQD systems
[\onlinecite{Ono}--\onlinecite{ZajacDM}], whose operation is
typically restricted to ultra-low temperatures due to the small
energy separation between bound and continuum states.
Nanoelectronic devices based on TDQDs therefore offer a promising
route toward robust high-temperature quantum functionality.

\begin{figure}[h]
\centering
\includegraphics[trim=1.cm 0cm 1.cm 0cm,clip,angle=0,scale=0.3]{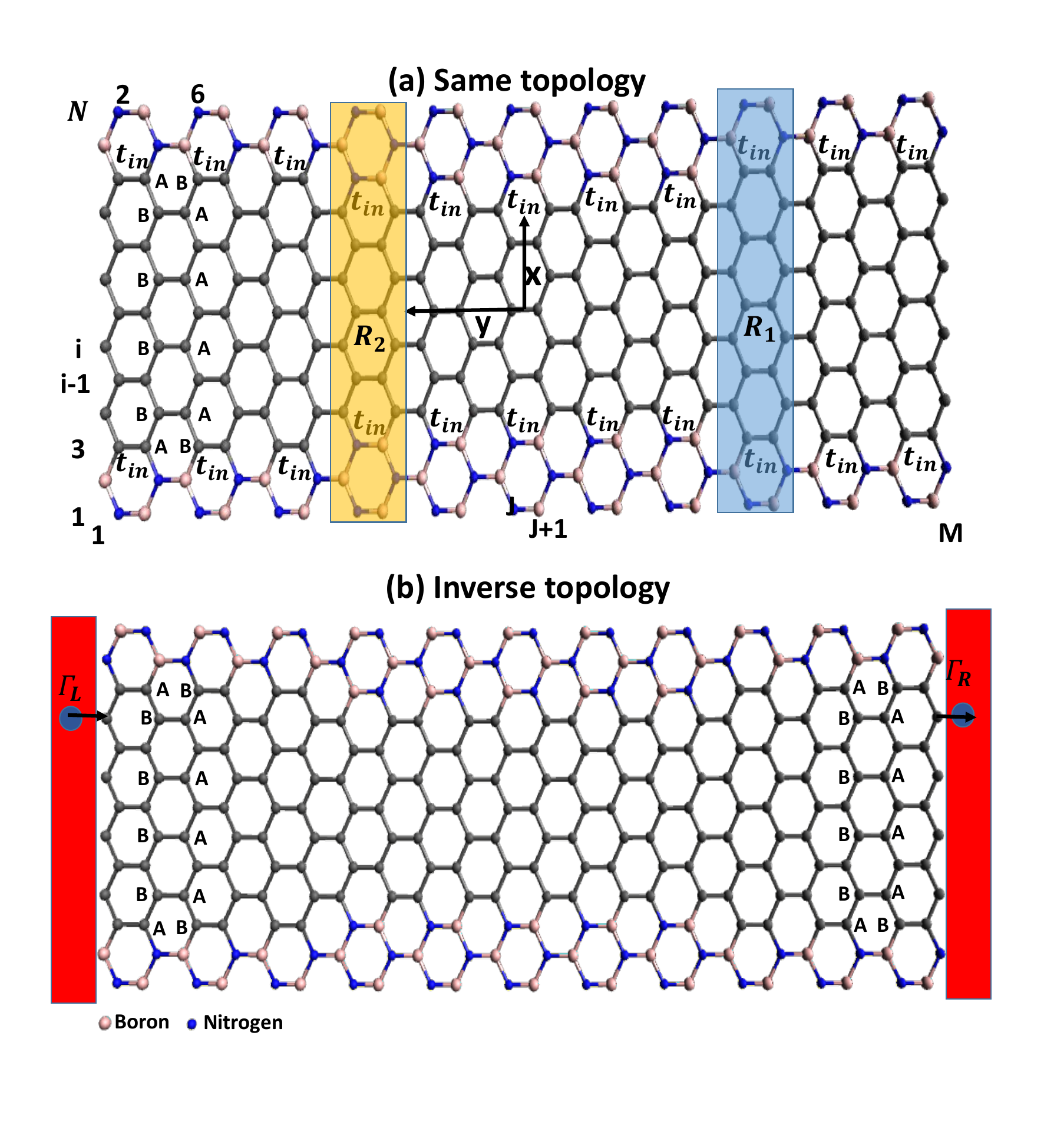}
\caption{Schematic illustration of an armchair graphene nanoribbon
heterostructure (AGNRH) laterally embedded in a hexagonal boron
nitride (BN) sheet, forming two distinct configurations: (a) a
same-topology structure, 2-BNNR/9-7-9-AGNRH/2-BNNR, and (b) a
reverse-topology structure, 2-BNNR/9-7-9-AGNRH/2-NBNR. White and
blue sites denote boron and nitrogen atoms, respectively.  $R_1$
and $R_2$ represent the unit cells (u.c.) of the
2-BNNR/9-AGNR/2-BNNR and 3-BNNR/7-AGNR/3-BNNR junctions,
respectively. The tunable parameter $t_{in}$ characterizes the
interfacial coupling between the AGNRH and the BNNRs. The symbols
$\Gamma_{L}$ ($\Gamma_{R}$) represent the electron tunneling rates
between the left (right) electrode and the leftmost (rightmost)
atoms at the zigzag edges.}
\end{figure}
\section{Calculation Methodology}
To investigate the transport properties of the BNNR/AGNRH/NBNR
heterojunction connected to metallic electrodes, we employ a
combination of a tight-binding model and the Green's function
technique [\onlinecite{QiaoZH}]. The system Hamiltonian,
illustrated in Fig.~1(b), is written as $H=H_0+H_{GNR}$, where
$H_0$ describes the electrodes and their coupling to the
nanoribbon system, and $H_{GNR}$ represents the BNNR/AGNRH/NBNR
structure.

\begin{small}
\begin{eqnarray}
H_0& = &\sum_{k} \epsilon_k a^{\dagger}_{k}a_{k}+
\sum_{k} \epsilon_k b^{\dagger}_{k}b_{k}\\
\nonumber &+&\sum_{\ell}\sum_{k}
V^L_{k,\ell,j}d^{\dagger}_{\ell,j}a_{k}
+\sum_{\ell}\sum_{k}V^R_{k,\ell,j}d^{\dagger}_{\ell,j}b_{k} + h.c.
\end{eqnarray}
\end{small}
The first two terms describe free electrons in the left and right
metallic electrodes. The operators $a^{\dagger}_{k}$
($b^{\dagger}_{k}$) create an electron with momentum $k$ and
energy $\epsilon_k$ in the left (right) electrode, respectively.
The coupling amplitudes $V^L_{k,\ell,j=1}$ and $V^R_{k,\ell,j=M}$
describe tunneling between the left (right) electrode and its
adjacent atom located in the $\ell$-th row at the zigzag edge.

\begin{small}
\begin{eqnarray}
H_{GNR}&= &\sum_{\ell,j} E_{\ell,j} d^{\dagger}_{\ell,j}d_{\ell,j}\\
\nonumber&-& \sum_{\ell,j}\sum_{\ell',j'} t_{(\ell,j),(\ell', j')}
d^{\dagger}_{\ell,j} d_{\ell',j'} + h.c,
\end{eqnarray}
\end{small}
Here, $E_{\ell,j}$ denotes the on-site energy of the orbital at
site $(\ell,j)$, while $d^{\dagger}_{\ell,j}$ and $d_{\ell,j}$ are
the electron creation and annihilation operators, respectively.
The parameter $t_{(\ell,j),(\ell', j')}$ characterizes the hopping
amplitude between sites $(\ell',j')$ and $(\ell,j)$. For the
BNNR/AGNRH/NBNR system, we assign on-site energies
$E_{B}=\Delta_B$, $E_{N}=-\Delta_N$, and $E_{C}=0$ for boron,
nitrogen, and carbon atoms, respectively. Variations in hopping
strengths between different atomic species are neglected, and the
condition $E_B=-E_N=\Delta = 2.7$~eV is adopted for simplicity
[\onlinecite{DingY}--\onlinecite{KuoJL}]. The nearest-neighbor
hopping energy is set to
$t_{(\ell,j),(\ell',j')}=t_{pp\pi}=2.7$~eV.

To quantify interfacial interactions induced by the BN
environment, we introduce an inter-nanoribbon hopping parameter
$t_{in}$ based on bulk boundary perturbation
approach[\onlinecite{DavidKuo}], as illustrated in Fig.~1(a). When
$t_{in}=0$, the BNNR/AGNRH/BNNR (or BNNR/AGNRH/NBNR) structure is
decoupled into three isolated segments--the lower BNNR, the
central 9-7-9-AGNRH, and the upper BNNR (or NBNR)--each subject to
vacuum boundary conditions. Varying $t_{in}$ effectively models
different strengths of lateral interface interactions arising from
surrounding two-dimensional insulating environments
[\onlinecite{PerepichkaDF}].

The bias-dependent transmission coefficient ${\cal
T}_{GNR}(\varepsilon)$ is calculated using ${\cal
T}_{GNR}(\varepsilon) =
4Tr[\Gamma_{L}(\varepsilon)G^{r}(\varepsilon)\Gamma_{R}(\varepsilon)G^{a}(\varepsilon)]$
[\onlinecite{DavidKuo1}], where $\Gamma_{L}(\varepsilon)$ and
$\Gamma_{R}(\varepsilon)$ denote the tunneling rate (in energy
units) at the left and right leads, respectively, and
$G^{r}(\varepsilon)$ and $G^{a}(\varepsilon)$ are the retarded and
advanced Green's functions of the GNRs, respectively. In terms of
tight-binding orbitals, $\Gamma_{\alpha}(\varepsilon)$ and Green's
functions are matrices. The expression for
$\Gamma_{L(R)}(\varepsilon)$ is derived from the imaginary part of
the self-energies, denoted as $\Sigma^r_{L(R)}(\varepsilon)$, and
is given by
$\Gamma_{L(R)}(\varepsilon)=-\text{Im}(\Sigma^r_{L(R)}(\varepsilon))=\pi\sum_k|V^{L(R)}_{k,\ell,j=1(M)}|^2\delta(\varepsilon-\epsilon_k)$.
In the wide-band limit, $\Gamma_{L(R)}(\varepsilon)$ is replaced
by an energy-independent constant matrix $\Gamma_{L(R)}$
[\onlinecite{DavidKuo}].

\section{Results and discussion}
Interface-induced modifications of AGNRs embedded in BN sheets
have been experimentally observed and shown to alter the band gaps
between conduction and valence subbands
[\onlinecite{Levendorf}--\onlinecite{GengDC}]. To elucidate how BN
environments affect the topological IFs of 9-7-9 -AGNRHs, we first
examine the influence of BNNRs with different topologies on
individual 7-AGNR and 9-AGNR with $R_1$ unit cells. Figures~2(a)
and 2(b) show the calculated band structures of
4-BNNR/7-AGNR/4-BNNR and 4-BNNR/9-AGNR/4-BNNR, corresponding to
same-topology configurations. The resulting band gaps are
$1.22$~eV and $1.19$~eV, respectively, compared with $1.0$~eV and
$0.9$~eV for isolated 7-AGNR and 9-AGNR under vacuum boundary
conditions.

For same-topology BN lateral embedding, both the lower and upper
BN edges induce sublattice potentials of the same sign.
Consequently, the boundary condition imposed by BN resembles that
of a vacuum (infinite potential barrier), except that the BN case
corresponds to a finite potential barrier. This explains why the
band-gap differences between vacuum and BN boundary conditions are
relatively small. Indeed, when $\Delta$ is set to a sufficiently
large value (e.g., $\Delta = 100~\mathrm{eV}$), the two results
become nearly indistinguishable. These findings indicate that BN
lateral embedding introduces only weak perturbations to the bulk
electronic structure when the surrounding BNNRs share the same
topology [\onlinecite{JungJ},\onlinecite{KuoJL}].

In contrast, for inverse-topology configurations shown in
Figs.~2(c) and 2(d), the band gaps of 4-BNNR/7-AGNR/4-NBNR and
4-BNNR/9-AGNR/4-NBNR are significantly reduced to $0.394$~eV and
$0.454$~eV, respectively. This pronounced gap renormalization
highlights the strong influence of the BN environment when the
lower BNNR has an opposite topology to the upper NBNR. In the
inverse-topology case, the sublattice potentials induced at the
lower and upper BN edges have opposite signs. Such a boundary
condition differs fundamentally from that of AGNRs under vacuum
boundary conditions, leading to substantial modifications of the
band-edge states.

These tight-binding results are in good agreement with previous
DFT calculations [\onlinecite{JungJ},\onlinecite{KuoJL}]. The
small residual discrepancies suggest that higher-order hopping
processes play a negligible role in determining the low-energy
conduction and valence subbands. Furthermore, we find that varying
the BNNR width ($n=6,8,10,\ldots$) does not qualitatively modify
the band-edge dispersions, thereby justifying the use of 4-BNNRs
in the following analysis.

\begin{figure}[h]
\centering
\includegraphics[angle=0,scale=0.3]{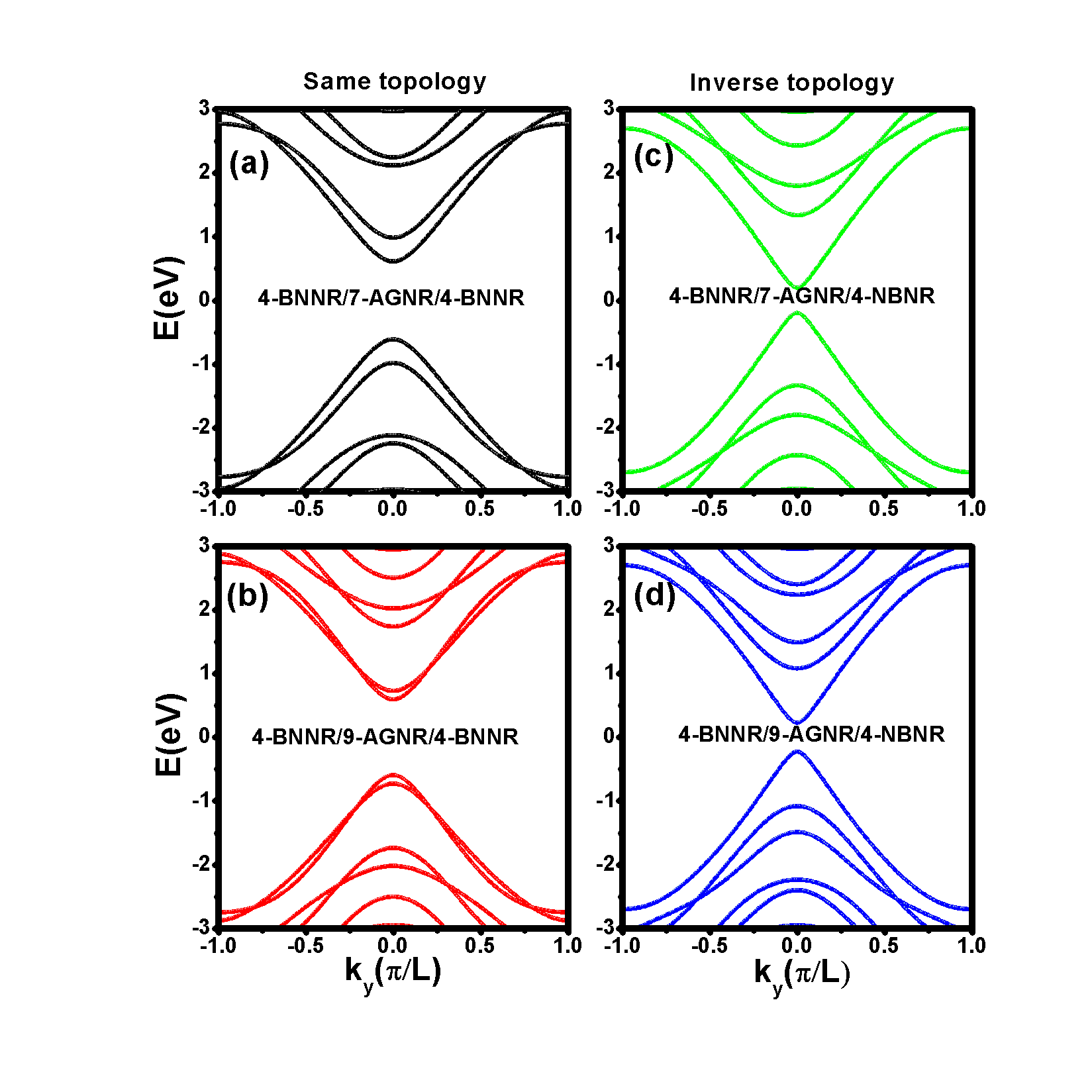}
\caption{Quasi-1D electronic band structures of AGNRs embedded in
hexagonal BN sheets. Same-topology configurations: (a)
4-BNNR/7-AGNR/4-BNNR and (b) 4-BNNR/9-AGNR/4-BNNR.
Inverse-topology configurations: (c) 4-BNNR/7-AGNR/4-NBNR and (d)
4-BNNR/9-AGNR/4-NBNR. We have adopted $t_{in}=2.7$~eV and $\Delta
= 2.7$~eV. $L$ is the length of $R_1$ unit cell.}
\end{figure}

In realistic experiments, AGNR and AGNRH segments possess finite
lengths, typically below $20$~nm[\onlinecite{DJRizzo}]. We
therefore examine the discrete energy spectra of finite 7-AGNR and
9-7-9 AGNRH segments embedded in BN environments with same and
inverse topology, as shown in Fig.~3. Figures~3(a) and 3(b)
display the energy levels of 4-BNNR/7-AGNR/4-BNNR and
4-BNNR/9-7-9-AGNRH/4-BNNR as functions of the interfacial hopping
strength $t_{in}$ for the same-topology case. Due to the finite
size of the 4-BNNR/7-AGNR/4-BNNR segment $(N=15,M=80)$, two in-gap
energy levels, denoted as $\Sigma_{ES,C}$ and $\Sigma_{ES,V}$,
emerge from the ESs, as shown in Fig.~3(a). In the absence of BN
coupling ($t_{in}=0$), these ESs form near-zero-energy modes. As
$t_{in}$ increases, $\Sigma_{ES,C}$ and $\Sigma_{ES,V}$ shift
markedly away from zero energy, whereas the bulk-derived states
remain largely unchanged. This behavior is consistent with the
infinite-length band structures shown in Fig.~2, indicating that
BN coupling primarily affects boundary-localized states in the
same-topology case.

Figure~3(b) shows four in-gap energy levels originating from two
ESs and two IFs in the 9-7-9 AGNRH segment. The IFs arise from the
effective end states of the central 7-AGNR segment with $R_2$ unit
cells, as discussed in our previous work [\onlinecite{DavidKuo}].
Within the physically relevant range $0.8~t\le t_{in}\le t$,
corresponding to realistic BN coupling strengths, the IF levels
$\Sigma_{IF,C(V)}$ are no longer well isolated from the bulk
states. This behavior contrasts sharply with the $t_{in}=0$ limit
and reflects the enhanced sublattice polarization induced by the
symmetric BN environment, which drives the IF levels toward the
bulk continuum.

The situation changes dramatically in the inverse-topology
configuration shown in Figs.~3(c) and 3(d), where finite 7-AGNR
and $9_6$-$7_8$-$9_6$ AGNRH segments are embedded between a BNNR
and an NBNR. Although the bulk band gap ($E_g = E_C - E_V$) is
substantially reduced for $0.8~t\le t_{in}\le t$, the energy
levels associated with the ESs and IFs in the 9-7-9 AGNRH remain
nearly unchanged and well separated from the bulk continuum. Such
robustness is highly nontrivial and indicates that the
inverse-topology configuration effectively preserves the IFs
against strong environmental perturbations.

\begin{figure}[h]
\centering
\includegraphics[angle=0,scale=0.3]{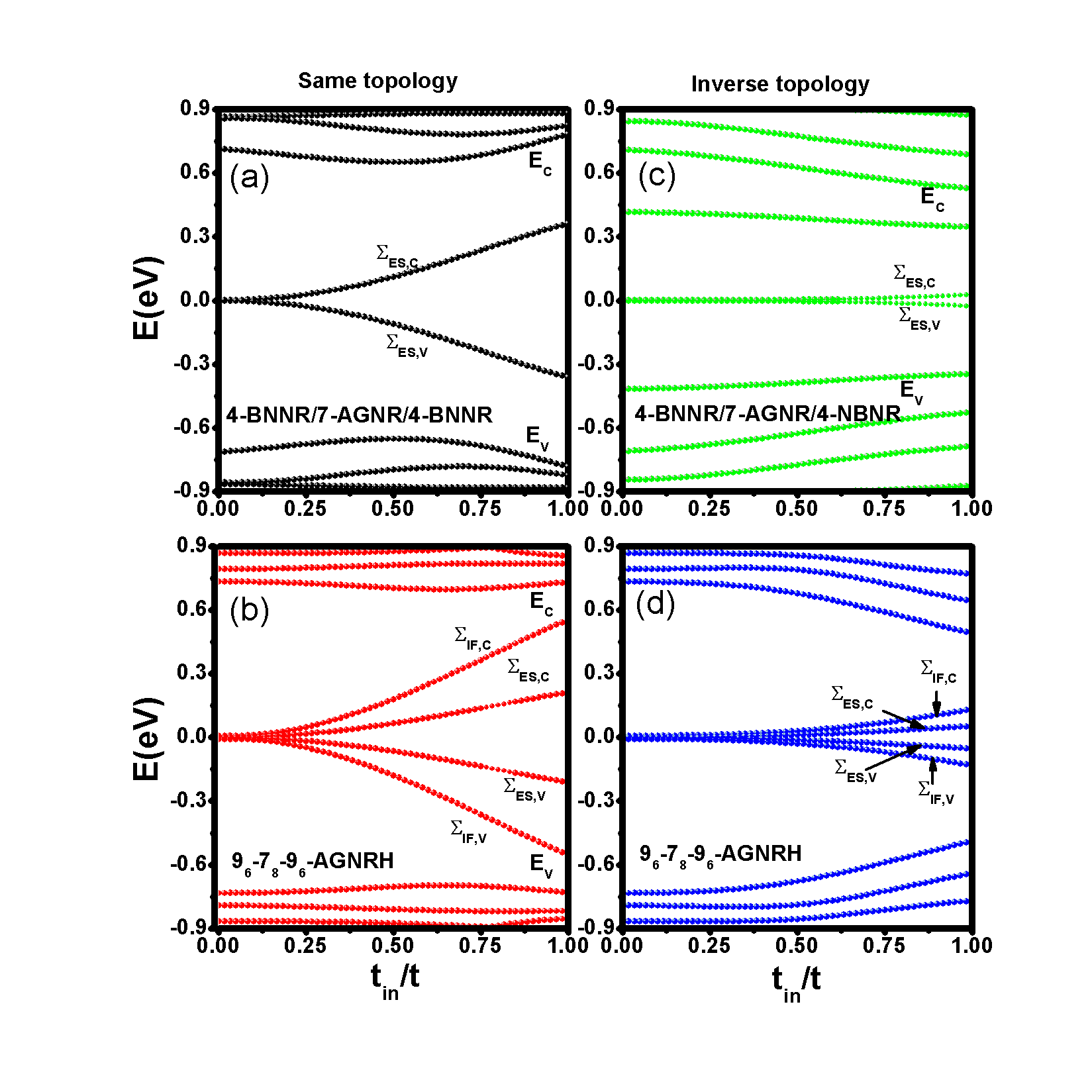}
\caption{Energy spectra of finite AGNR (AGNRH) segments as
functions of interfacial hopping $t_{in}$ at $\Delta = 2.7$~eV.
Same-topology scenario: (a) 4-BNNR/7-AGNR/4-BNNR and (b)
4-BNNR/9-7-9-AGNRH/4-BNNR. Inverse-topology scenario: (c)
4-BNNR/7-AGNR/4-NBNR and (d) 4-BNNR/9-7-9-AGNRH/4-NBNR.}
\end{figure}

To clarify the contrasting behaviors observed in Figs.~3(b) and
3(d), we plot the charge densities of $\Sigma_{IF,C}$ for
different $t_{in}$ values in Fig.~4. Figures~4(a)-4(c) and
Figs.~4(d)-4(f) correspond to the inverse-topology and
same-topology configurations, respectively. For $t_{in} = 0$,
shown in Fig.~4(a), the AGNRH is completely decoupled from the BN
environment, corresponding to vacuum boundary conditions. The
state $\Sigma_{IF,C}$ originates from the coupling between a 9-7
junction IF state with A-sublattice chirality
($\Psi_{IF,A}(t_{in}=0)$) and a 7-9 junction IF state with
B-sublattice chirality ($\Psi_{IF,B}(t_{in}=0)$). When a weak
coupling $t_{in} = 0.2~t$ is introduced, the charge density of
$\Sigma_{IF,C}$ partially penetrates into the BN regions
[Fig.~4(b)] while preserving mirror symmetry across the ribbon.
Even for strong coupling $t_{in} = 0.8~t$, as shown in Fig.~4(c),
$\Sigma_{IF,C}$ maintains nearly equal contributions from
$\psi_{IF,A}(t_{in}\neq 0)$ and $\psi_{IF,B}(t_{in}\neq 0)$. The
charge-density distribution remains mirror-inverted across the
ribbon due to the global inversion symmetry of the system. The
increased energy separation between $\Sigma_{IF,C}$ and
$\Sigma_{IF,V}$ is mainly attributed to the enhanced overlap
between the wave functions $\psi_{IF,A}$ and $\psi_{IF,B}$. For
finite $t_{in}$, the decay lengths of $\psi_{IF,A}$ and
$\psi_{IF,B}$ increase as a result of interfacial-induced
delocalization. Notably, charge densities simultaneously appear on
both A- and B-sublattice sites at rows $i = 5$ and $i = 13$,
reflecting the symmetric hybridization between the two junction
states.

In contrast, Figs.~4(d)-4(f) demonstrate that in the same-topology
configuration the IF charge densities become strongly polarized
even at $t_{in} = 0.2~t$, with dominant contributions from
$\psi_{IF,A}$. The A(B)-sublattice IF states couple to nitride
(boron) atoms at the lower and upper BN edges, respectively,
leading to the formation of new bonding and antibonding states
[\onlinecite{PrunedaJM}]. As a result, $\Sigma_{IF,C(V)}$ shifts
above (below) the mid-gap energy. This behavior indicates that the
symmetric BN environment generates an effective field along the
armchair direction, thereby breaking the inversion symmetry
between the 9-7 and 7-9 junctions.

To further elucidate the influence of BN sheets on the IFs shown
in Figs.~3 and 4, we introduce an effective Hamiltonian for the IF
subspace:

\begin{small}
\begin{eqnarray}
&&H_{eff,IF}\\
&=&\left[
\begin{array}{cc}
E_L+\Sigma_A & t_{eff,LR}\\
t_{eff,LR} & E_R+\Sigma_B\\
\end{array}\right]\nonumber,
\end{eqnarray}
\end{small}

where $E_L = 0$ and $E_R = 0$ denote the energy levels of the 9-7
and 7-9 junction IF states in the absence of $t_{in}$. $\Sigma_A$
and $\Sigma_B$ represent the self-energies induced by the
interfacial hopping $t_{in}$.

For the inverse-topology configuration, the effective Hamiltonian
reduces to

\begin{small}
\begin{eqnarray}
&&H_{eff,IF,I}\\
&=&\left[
\begin{array}{cc}
0 & t_{eff,LR}\\
t_{eff,LR} & 0\\
\end{array}\right]\nonumber,
\end{eqnarray}
\end{small}

since the self-energies satisfy $\Sigma_{A(B)} = 0$. More
specifically, $\Sigma_{A(B)} = t_{in}^2(G_{Bo}+G_{Ni})$, where
$G_{Bo}$ and $G_{Ni}$ are the Green's functions describing
electron propagation on boron and nitride atoms, respectively.
Under the approximation $\Delta_B = -\Delta_N$, one has $G_{Bo}
\approx -G_{Ni}$, leading to nearly vanishing self-energies. In
the inverse-topology configuration, the upper BN ribbon is flipped
relative to the lower one, reversing the BN sublattice alignment
at the two interfaces and restoring chiral symmetry in the IF
subspace.

For the same-topology configuration, the effective Hamiltonian
becomes

\begin{small}
\begin{eqnarray}
&&H_{eff,IF,S}\\
&=&\left[
\begin{array}{cc}
2\Sigma_{Ni} & t_{eff,LR}\\
t_{eff,LR} & 2\Sigma_{Bo}\\
\end{array}\right]\nonumber.
\end{eqnarray}
\end{small}

In this case, the self-energies satisfy $\Sigma_{A(B)} =
2\Sigma_{Ni(Bo)}$, which significantly breaks the chiral symmetry
of the IF states. For the same-topology configuration, A sites in
graphene couple to the same BN species (nitride atoms) on both
sides, while B sites couple to boron atoms on both sides. The
resulting asymmetric on-site renormalization explains the
polarization of the IF charge densities observed in Fig.~4. The
eigenstates of $H_{eff,IF,I}$ and $H_{eff,IF,S}$ therefore provide
a consistent interpretation of the numerical charge-density
distributions.

\begin{figure}[h]
\centering
\includegraphics[angle=0,scale=0.2]{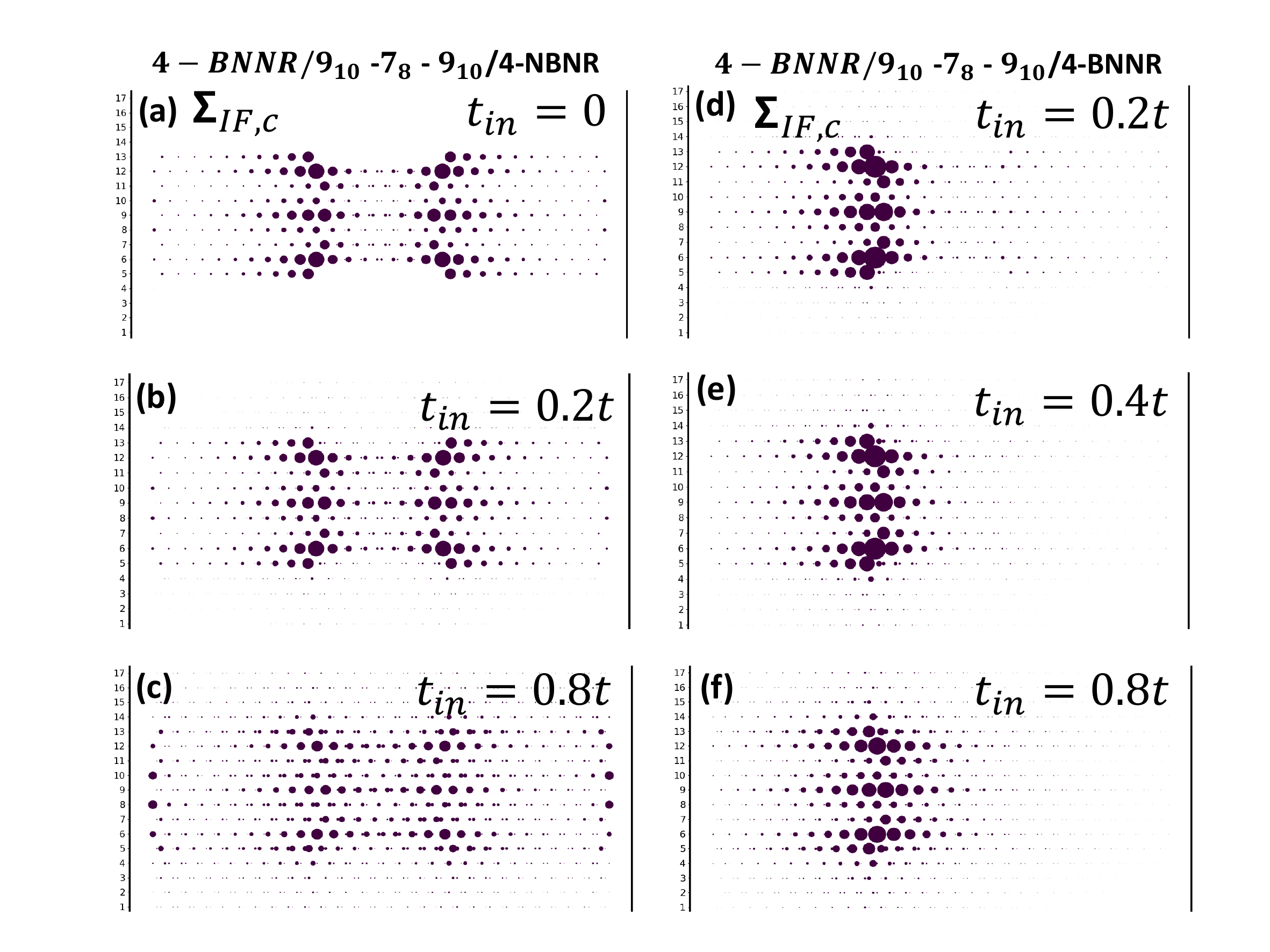}
\caption{Charge densities of the interface state $\Sigma_{IF,C}$
in $9_{10}-7_8-9_{10}$-AGNRH embedded in BN at $\Delta = 2.7$~eV.
(a)-(c) Inverse-topology: $t_{in}=0$ ($\Sigma_{IF,C}=7.53$~meV),
$t_{in}=0.2~t$ ($\Sigma_{IF,C}=8.81$~meV), $t_{in}=0.8~t$
($\Sigma_{IF,C}=50.96$~meV). (d)-(f) Same-topology: $t_{in}=0.2~t$
($\Sigma_{IF,C}=31.35$~meV), $t_{in}=0.4~t$
($\Sigma_{IF,C}=117.57$~meV), $t_{in}=0.8~t$
($\Sigma_{IF,C}=0.402$~eV).}
\end{figure}

Finally, we examine the transport consequences of the robust IFs
in inverse-topology structures. Figure~5 presents the calculated
transmission coefficient ${\cal T}_{GNR}(\varepsilon)$ for
4-BNNR/$9_6$-$7_8$-$9_6$-AGNRH/4-NBNR at various values of
$t_{in}$. Pronounced transmission peaks located at $\Sigma_{IF,C}$
and $\Sigma_{IF,V}$ confirm that the system effectively operates
as a TDQD. In contrast, no distinct transport signatures are
observed at the ES-related energy levels, consistent with their
short decay lengths relative to the channel length ($M = 80$),
which suppresses their coupling to the electrodes
[\onlinecite{DavidKuo3},\onlinecite{DavidKuo4}].

The transmission spectra in the vicinity of the IF resonances are
well described by ${\cal T}_{TDQD}(\varepsilon)=
\frac{4\Gamma_{e,L}t^2_{eff,LR}\Gamma_{e,R}}{|(\varepsilon-E_L+i\Gamma_{e,L})(\varepsilon-E_R+i\Gamma_{e,R})-t^2_{eff,LR}|^2}
$ in the absence of inter-dot and intra-dot Coulomb interactions
[\onlinecite{DavidKuo2}]. Here, $\Gamma_{e,L(R)}$ denotes the
effective tunneling rate between the electrodes and the left
(right) TQD, $E_{L(R)} = 0$ represent the on-site energies of the
two IF states, and $t_{eff,LR}$ is the effective interdot hopping
strength. The zero-temperature electrical conductance is given by
$G_e(\mu)=(2e^2/h){\cal T}_{TDQD}(\varepsilon=\mu)$, which reaches
the conductance quantum $G_0=2e^2/h$ when $\mu$ aligns with the
energy levels $\Sigma_{IF,C}$ and $\Sigma_{IF,V}$. Here, $\mu$
denotes the Fermi energy of the electrodes.

Notably, the spectra of $G_e(\mu)$ indicate that the effective
hopping $t_{eff,LR}$ between the IFs in inverse-topology BN
environments is enhanced compared with the vacuum boundary
condition. This enhancement underscores the constructive role of
topology-engineered environments in stabilizing the IFs and
strengthening coherent topological transport. In Appendix~A, we
further investigate the electronic properties of a 15-13-15-AGNRH
segment embedded in BN sheets, demonstrating that the above
conclusions are not restricted to the 9-7-9 geometry.

\begin{figure}[h]
\centering
\includegraphics[angle=0,scale=0.3]{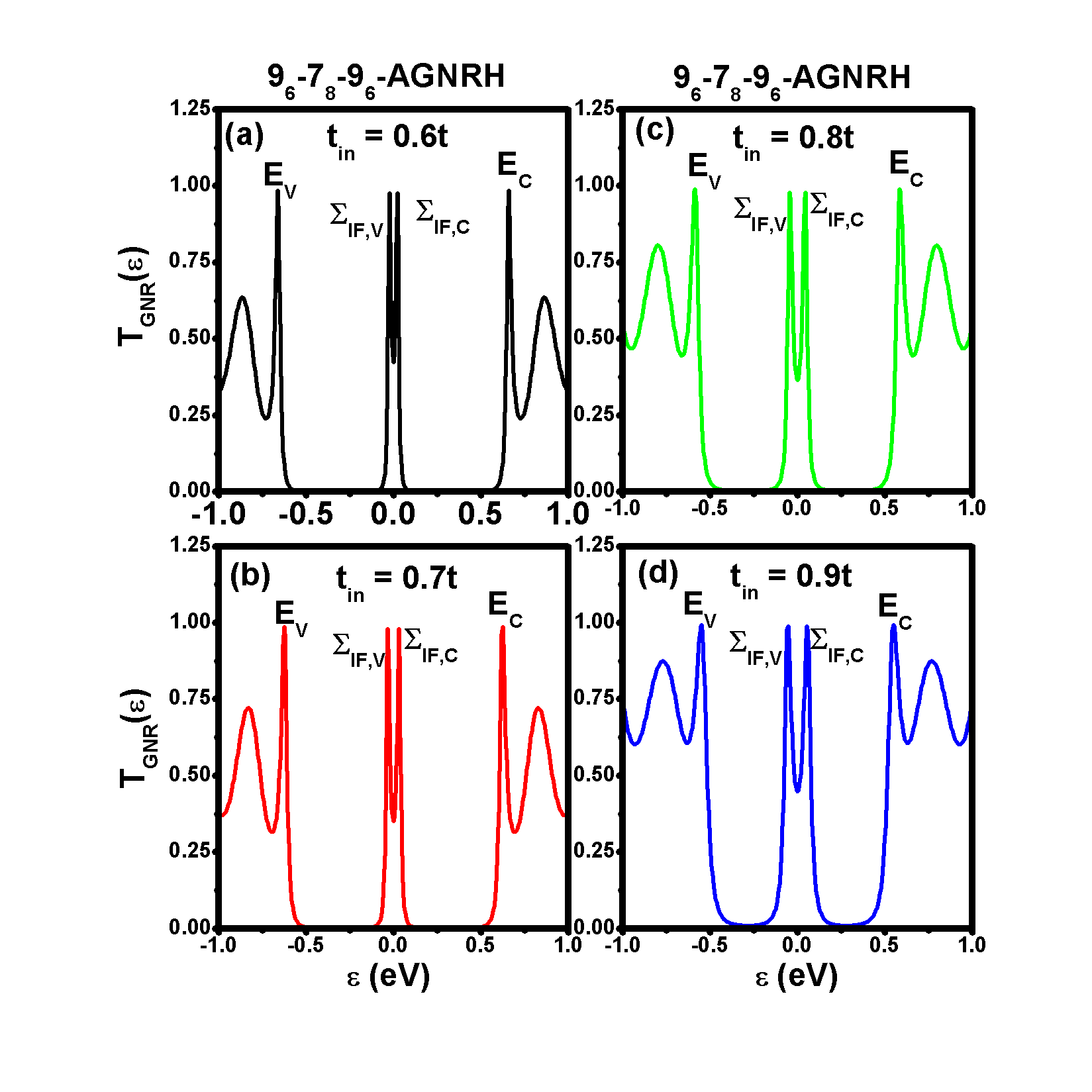}
\caption{Transmission coefficients ${\cal T}_{GNR}(\varepsilon)$
of the 4-BNNR/$9_6-7_8-9_6$-AGNRH/4-NBNR structure for various
$t_{in}$ at $\Gamma_L=\Gamma_R=\Gamma_t = 2.7$~eV. (a) $t_{in} =
0.6~t$, (b) $t_{in} = 0.7~t$, (c) $t_{in} = 0.8~t$, and (d)
$t_{in} = 0.9~t$.}
\end{figure}

\section{Conclusion}
In this work, we theoretically investigated the electronic and
transport properties of 9-7-9 and 15-13-15 AGNRHs laterally
embedded in boron nitride (BN) sheets. Two sandwich configurations
were considered: $n$-BNNR/AGNRH/$n$-BNNR and
$n$-BNNR/AGNRH/$n$-NBNR, corresponding to same-topology and
reverse-topology environments, respectively. These distinct
topological configurations give rise to fundamentally different
behaviors of the ESs and IFs in the AGNRH.

Within a bulk-boundary perturbation framework, we demonstrate that
the topology of the surrounding BN nanoribbons plays a decisive
role in determining the stability of the IFs. For same-topology BN
environments, interfacial coupling between BN and C atoms induces
asymmetric self-energies on the two sublattices. This effect
strongly lifts the energy level of the IF state with A-chirality
relative to that with B-chirality, breaking the chiral symmetry of
the IF subspace and destroying their functionality as TDQDs. As a
consequence, the IF levels hybridize with bulk states and lose
their spectral isolation.

In contrast, when the surrounding BN nanoribbons realize a
reverse-topology configuration, the global inversion symmetry is
restored through mirror-inverted interfaces. In this case, the
self-energy corrections from the BN environment largely cancel,
leaving the IF energy levels only weakly perturbed. The IFs in
$n$-BNNR/AGNRH/$n$-NBNR therefore remain well separated from the
bulk continuum and retain their TDQD functionality, as confirmed
by robust resonant transport reaching the conductance quantum.

Our results establish topology-engineered BN environments as an
effective route for stabilizing and even enhancing topological
interface states in AGNRHs. The demonstrated symmetry-preserved
IFs provide a promising platform for realizing robust quantum
functionalities in graphene-based nanoelectronic devices,
including high-temperature quantum-dot
operation[\onlinecite{ChouSY},\onlinecite{ChouSY1}] and qubit-like
applications[\onlinecite{LossD}--\onlinecite{Davidkuo5}].


{}

\textbf{Data availability}\\

The data presented in this study are available upon reasonable
request.

\textbf{Conflicts of interest}\\
There are no conflicts to declare

 \numberwithin{figure}{section} \numberwithin{equation}{section}

\setcounter{section}{0}
\setcounter{equation}{0} 

\mbox{}\\

\appendix
\numberwithin{figure}{section}
\numberwithin{equation}{section}

\section{Wider width AGNRHs embedded into boron nitride sheet}
In the main text, the electronic properties of two sandwich
structures, $n$-BNNR/9-7-9-AGNRH/$n$-BNNR and
$n$-BNNR/9-7-9-AGNRH/$n$-NBNR, were calculated and analyzed. The
former corresponds to a same-topology configuration in which the
lower and upper BNNRs share identical topology, whereas the latter
represents a reverse-topology configuration. Although only
9-7-9-AGNRH segments have been experimentally synthesized
[\onlinecite{DJRizzo}], in this appendix we investigate the
environmental effects on the transport properties of
15-13-15-AGNRHs in order to examine the size dependence of AGNRH
responses to BN embedding.

Figure~A.1 presents the calculated energy levels of
4-BNNR/15-13-15-AGNRH/4-BNNR and 4-BNNR/15-13-15-AGNRH/4-NBNR
segments as functions of the interfacial hopping strength
$t_{in}$. The overall trends for the 15-13-15-AGNRH in Fig.~A.1
closely resemble those of the 9-7-9 AGNRH shown in Figs.~3(b) and
3(d). However, the bulk band gap of 15-13-15-AGNRH is smaller than
that of 9-7-9-AGNRH in both the same-topology and reverse-topology
configurations.

In 15-13-15-AGNRHs, six in-gap energy levels appear, consisting of
four ESs and two IFs. This follows from the fact that each 15-AGNR
segment hosts two left ESs with A-sublattice chirality and two
right ESs with B-sublattice chirality. The number of ESs for an
$N$-AGNR segment with $R_1$ unit cells is given by
$N_{es}=(N-3)/6$ and $N_{es}=(N-1)/6$ for $3p$-width and
$(3p+1)$-width AGNRs, respectively. For AGNR segments with $R_2$
unit cells, the number of ESs becomes $N_{es}+1$. Consequently,
the central 13-AGNR segment possesses three ESs at each terminus.
The number of IFs at each junction is determined by
$N_{IF}=|N_{es,cen}-N_{es,out}|$, where $N_{es,cen}$ and
$N_{es,out}$ denote the numbers of ESs in the central and outer
AGNR segments, respectively [\onlinecite{DavidKuo}]. At the 15-13
junction, one IF state with A-chirality emerges, whereas at the
13-15 junction one IF state with B-chirality appears.

In the same topology configuration shown in Fig. A. 1(a), the
energy-level separation $\Delta_{IF}$ between $\Sigma_{IF,C}$ and
$\Sigma_{IF,V}$ for $t_{in} = 0$ originates from the wave-function
overlap between the A-chirality and B-chirality IFs . We note that
$\Delta_{IF}$ in 15-13-15-AGNRH segments is substantially larger
than that in 9-7-9-AGNRH segments, reflecting stronger intrinsic
coupling between the two IFs. When $t_{in} \le 0.1~t$,
$\Delta_{IF}$ remains nearly unchanged, whereas for $t_{in} >
0.2~t$, $\Delta_{IF}$ exhibits an approximate $t_{in}^2$
dependence. This behavior is attributed to the gradual decoupling
between the A-chirality and B-chirality IFs as their spatial decay
lengths are modified (see Fig.~A.2(c)).

In the reverse-topology configuration shown in Fig.~A.1(b),
$\Delta_{IF}=2t_{eff,LR}$ remains nearly constant for $t_{in} \le
0.3~t$. At $t_{in} = t$, we obtain $E_g = 0.9$~eV and $\Delta_{IF}
= 0.362$~eV for the $15_6$-$13_8$-$15_6$-AGNRH segment, compared
with $E_g = 0.978$~eV and $\Delta_{IF} = 0.264$~eV for the
$9_6$-$7_8$-$9_6$-AGNRH segment. According to Eq.~(4), the
variation of $\Delta_{IF}$ with respect to $t_{in}$ is governed by
the effective inter-IF coupling $t_{eff,LR}$. The results of
Fig.~A.1(b) therefore indicate that the coherent length of IFs in
15-13-15-AGNRH segments is longer than that in 9-7-9-AGNRH
segments, even in the presence of BN embedding.

\begin{figure}[h]
\centering
\includegraphics[angle=0,scale=0.3]{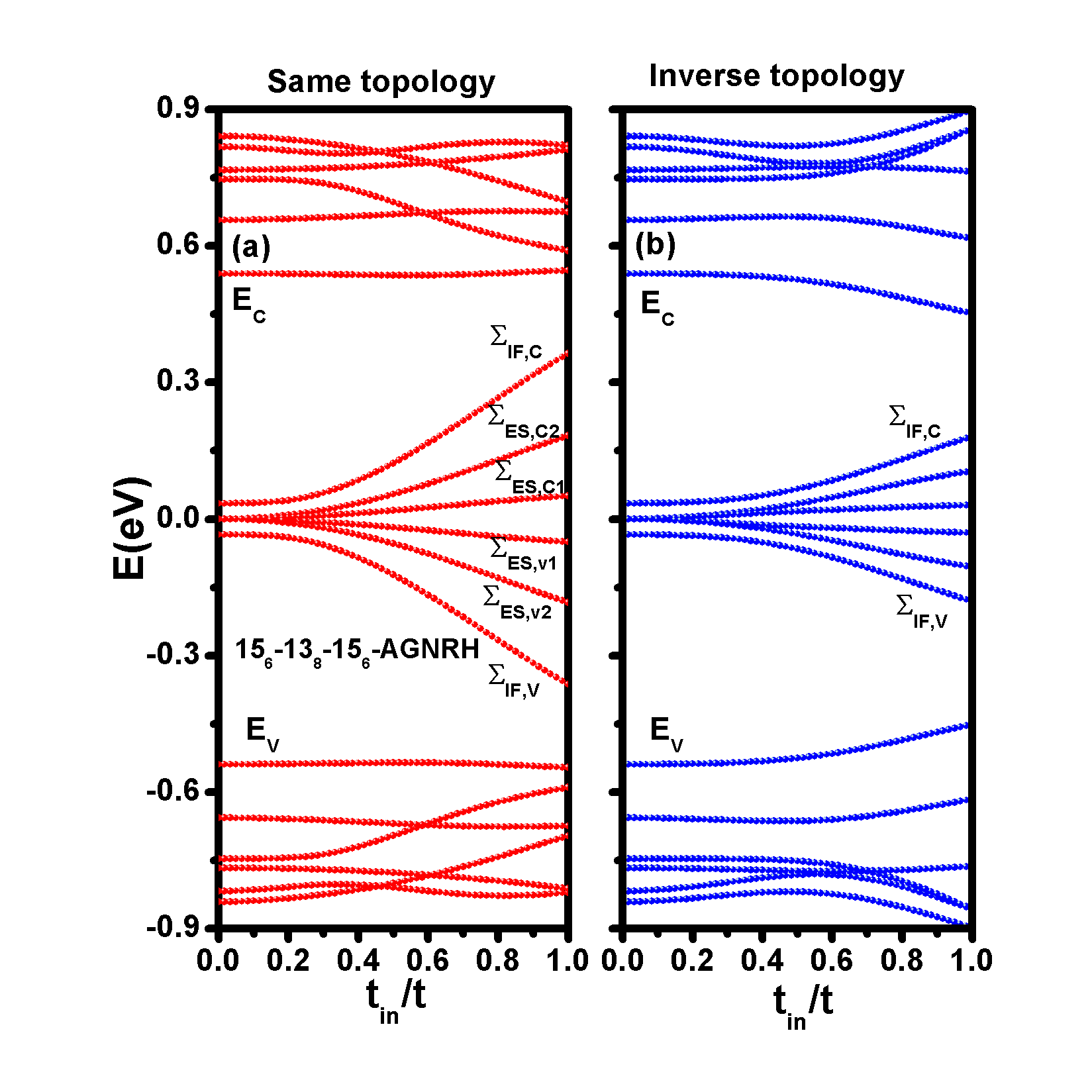}
\caption{Energy spectra of finite $15_6$-$13_8$-$15_6$ AGNRH
segments as a function of the interfacial hopping $t_{in}$ at
$\Delta = 2.7$~eV. (a) Same-topology configuration:
4-BNNR/$15_6$-$13_8$-$15_6$-AGNRH/4-BNNR. (b) Inverse-topology
configuration: 4-BNNR/$15_6$-$13_8$-$15_6$-AGNRH/4-NBNR. The
spectra illustrate the evolution of end states (ESs) and interface
states (IFs) under increasing BN coupling.}
\end{figure}

In Fig.~A.1(a), the energy levels of the topological states are
shifted due to the polarization induced by the BN sheets. To
further elucidate the polarization effects arising from BNNRs in
the same-topology scenario, we plot the charge densities of
$\Sigma_{IF,C}$ for the $15_6$-$13_8$-$15_6$ AGNRH at different
interfacial hopping strengths $t_{in}$ in Fig.~A.2.
Figures~A.2(a)-2(f) correspond to $t_{in}=0$, $t_{in}=0.2~t$,
$t_{in}=0.4~t$, $t_{in}=0.6~t$, $t_{in}=0.8~t$, and $t_{in}=t$,
respectively. For $t_{in}=0$ and $t_{in}=0.2~t$, the charge
densities of $\Sigma_{IF,C}$ are mostly confined within the
graphene ribbon. However, even at $t_{in}=0.2~t$, the A-sublattice
sites exhibit a noticeable enhancement of charge density,
indicating the onset of BN-induced polarization. When
$t_{in}=0.4~t$, the charge density becomes predominantly localized
on the A-sublattice sites, which reflects the significant
self-energy contributions $\Sigma_{Ni}$ and $\Sigma_{Bo}$ in
Eq.~(5).

As $t_{in}$ increases further to $0.6~t$, $0.8~t$, and $t$, the
charge density of $\Sigma_{IF,C}$ extends to the B-sublattice
sites at the BN edges (corresponding to nitride atom positions).
Simultaneously, the B-sublattice sites at rows $i = 5$ and $i =
19$ in the left 15-AGNR segment also display enhanced charge
density. This indicates a strong coupling between the A- and
B-sublattice sites within the left 15-AGNR segment, resulting in a
wave-function delocalization of $\Sigma_{IF,C}$ across the ribbon
and into the BN edges.

\begin{figure}[h]
\centering
\includegraphics[angle=0,scale=0.2]{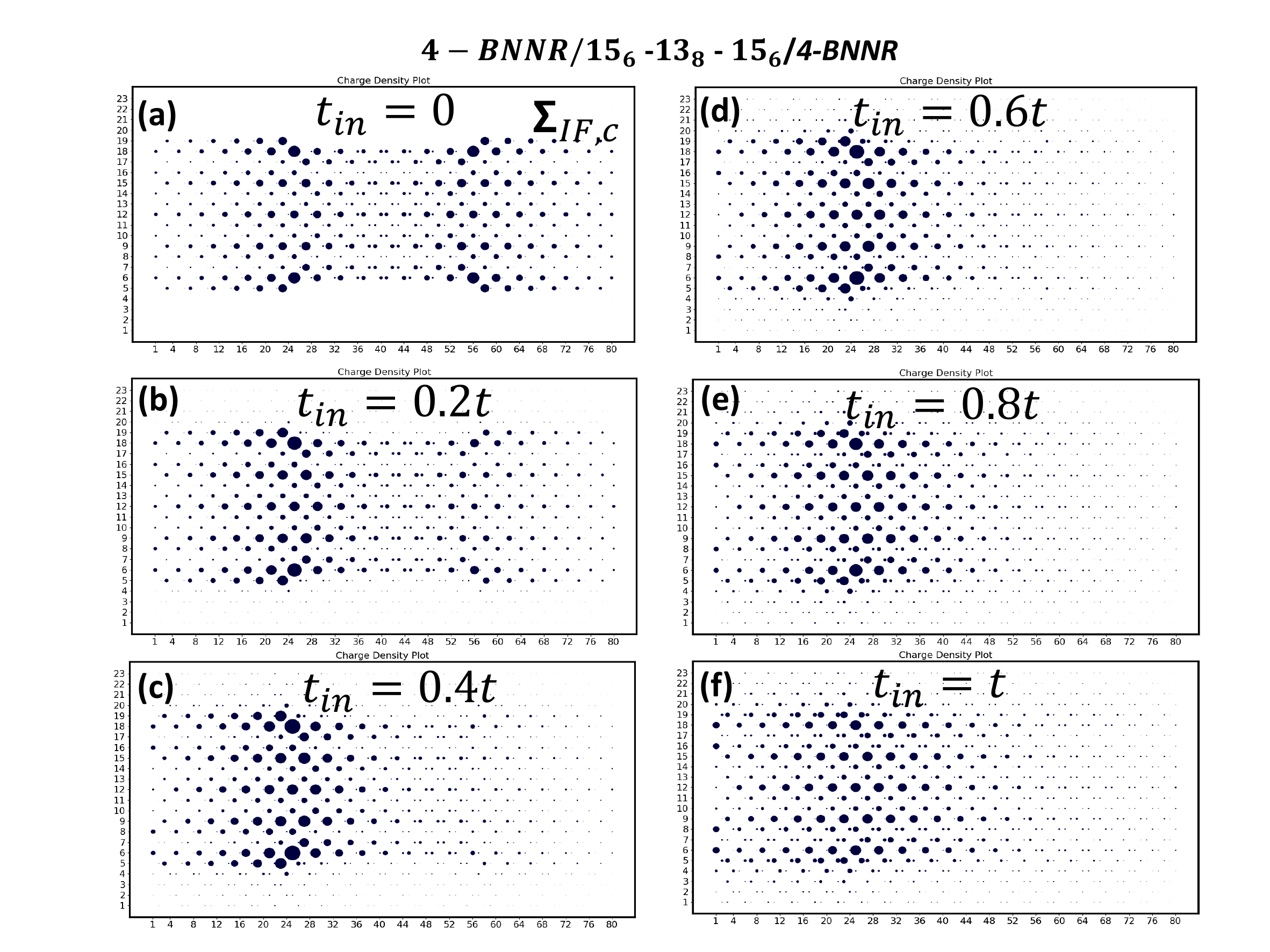}
\caption{Charge density distributions of the interface state
$\Sigma_{IF,C}$ in the 4-BNNR/$15_6$-$13_8$-$15_6$-AGNRH/4-BNNR
segment at $\Delta = 2.7$~eV. Panels (a)-(f) correspond to the
same-topology configuration with interfacial hopping $t_{in}=0$
($\Sigma_{IF,C}=34.27$~meV), $0.2~t$ ($\Sigma_{IF,C}=40.9$~meV),
$0.4~t$ ($\Sigma_{IF,C}=85.2$~meV), $0.6~t$
($\Sigma_{IF,C}=0.167$~eV), $0.8~t$ ($\Sigma_{IF,C}=0.266$~eV),
and $t$ ($\Sigma_{IF,C}=0.363$~eV), illustrating the progressive
polarization and delocalization of the interface state induced by
BN coupling.}
\end{figure}

To examine the transport properties of the
4-BNNR/15-13-15-AGNRH/4-BNNR segment, we calculate the
transmission coefficient ${\cal T}_{GNR}(\varepsilon)$ as a
function of applied voltage, which generates a uniform electric
field along the negative $y$-direction, as shown in Fig.~A.3 for
interfacial hopping $t_{in}=0.8~t$
[\onlinecite{DingY}--\onlinecite{GSSeal}]. The potential induced
by the electric field is incorporated into the system Hamiltonian
(Eq.~(2)) as $U_y = -eF_y y$, where $F_y = V_y/L_a$, $V_y$ is the
applied bias, and $L_a$ is the length of the AGNRH
segment[\onlinecite{DavidKuo3},\onlinecite{DavidKuo4}].

In Fig.~A. 3(a) at $V_y = 0$, two small transmission peaks labeled
$\Sigma_{IF,C}$ and $\Sigma_{IF,V}$ appear within the bulk gap.
Charge transport through these IF levels can be effectively
modeled as a single quantum dot (QD) with one energy level and
asymmetric tunneling rates $\Gamma_{e,L} \neq \Gamma_{e,R}$
[\onlinecite{Davidkuo5}]. This demonstrates that the IFs in
4-BNNR/15-13-15-AGNRH/4-BNNR behave as a single TQD at zero bias.
Increasing the tunneling rates between the zigzag edge structures
of the AGNRH and the electrodes, from $\Gamma_{t} =
0.56~\text{eV}$ to $2.7~\text{eV}$, enhances the peak amplitudes,
although ${\cal T}_{GNR}$ remains relatively small due to the
asymmetry of $\Gamma_{e,L}$ and $\Gamma_{e,R}$, as suggested by
the charge density distribution in Fig.~A.2(e). As illustrated in
Figs.~A.3(b)-(d), the separation between the $\Sigma_{IF,C}$ and
$\Sigma_{IF,V}$ peaks decreases as $V_y$ increases. Notably,
${\cal T}_{GNR}$ reaches unity at $V_y = 1.215$~V with $\Gamma_t =
0.56~\text{eV}$ in Fig.~A.3(d). These results indicate that the
4-BNNR/15-13-15-AGNRH/4-BNNR segment undergoes a transition from a
single TQD to a TDQD under electric-field modulation. Finally, we
would like to point out that our conclusions remain valid even in
the case of $\Delta_B \neq -\Delta_{N}$. For example, when the
energy level of boron atoms is $\Delta_B = 2.329~eV$ and the
energy level of nitride atoms is $\Delta_N = -2.499~eV$, the band
inversion symmetry shown in Fig. A.1 is slightly lifted. However,
the transmission coefficient still exhibits the resonant tunneling
characteristic associated with topological interface states in the
4-BNNR/AGNRH/4-NBNR configuration.

\begin{figure}[h]
\centering
\includegraphics[angle=0,scale=0.3]{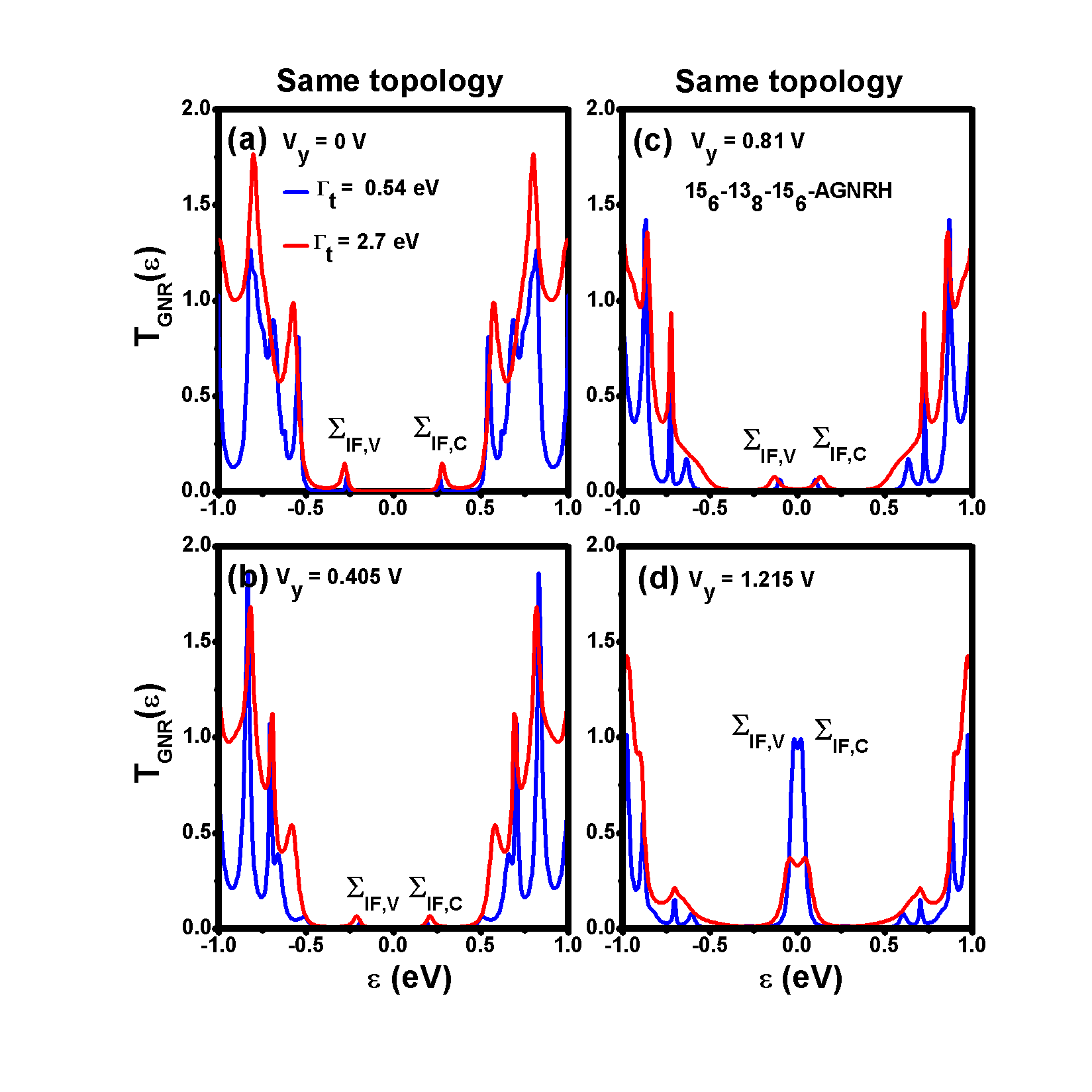}
\caption{Transmission coefficients ${\cal T}_{GNR}(\varepsilon)$
of the 4-BNNR/$15_6$-$13_8$-$15_6$-AGNRH/4-BNNR segment at
$t_{in}=0.8~t$ under various applied voltages $V_y$: (a) $V_y=0$,
(b) $V_y=0.405$~V, (c) $V_y=0.81$~V, and (d) $V_y=1.215$~V. Blue
and red curves correspond to tunneling rates
$\Gamma_L=\Gamma_R=\Gamma_t = 0.56$~eV and $\Gamma_t = 2.7$~eV,
respectively. The panels illustrate the evolution from a single
topological quantum dot at zero bias to a topological double
quantum dot under electric-field modulation.}
\end{figure}

{\bf Acknowledgments}\\
This work was supported by the National Science and Technology
Council, Taiwan under Contract No. MOST 107-2112-M-008-023MY2.

\mbox{}\\
E-mail address: mtkuo@ee.ncu.edu.tw\\




\end{document}